\documentclass[aps,prl,twocolumn,superscriptaddress]{revtex4-1}
\usepackage{bm}
\usepackage{graphicx}
\usepackage{color}
\usepackage{braket}
\usepackage{amsmath,amssymb,amsfonts,amsthm,mathtools}
\usepackage{enumerate}
\usepackage{enumitem}
\usepackage[colorlinks=true,linkcolor=blue,anchorcolor=red,citecolor=blue,urlcolor=blue]{hyperref}
\usepackage{titlesec}
\usepackage{tikz-cd}
\usepackage{float}
\usepackage{multirow}
\usepackage[title]{appendix}
\usepackage{pdfpages}
\usepackage{changes}
\usepackage{stackengine}
\usepackage{makecell}
\usepackage{caption}
\usepackage{pdfpages}

\makeatletter
\AtBeginDocument{\let\LS@rot\@undefined}
\makeatother

\makeatletter 
\renewcommand\NAT@biblabelnum[1]{#1.} 
\makeatother

\begin{document}

\title{Projective Spacetime Symmetry of Spacetime Crystals}

\author{Zheng Zhang}
\affiliation{National Laboratory of Solid State Microstructures and Department of Physics, Nanjing University, Nanjing 210093, China}
\affiliation{The University of HongKong Shenzhen Institute of Research and Innovation, Shenzhen 518057, China}

\author{Z. Y. Chen}
\affiliation{National Laboratory of Solid State Microstructures and Department of Physics, Nanjing University, Nanjing 210093, China}
\affiliation{The University of HongKong Shenzhen Institute of Research and Innovation, Shenzhen 518057, China}

\author{Y. X. Zhao}
\email[]{yuxinphy@hku.hk}

\affiliation{Department of Physics and HKU-UCAS Joint Institute for Theoretical
	and Computational Physics at Hong Kong, The University of Hong Kong,
	Pokfulam Road, Hong Kong, China}
\affiliation{HK Institute of Quantum Science \& Technology, The University of Hong Kong,
	Pokfulam Road, Hong Kong, China}

\begin{abstract}
Wigner's seminal work on the Poincar\'{e} group revealed one of the fundamental principles of quantum theory: symmetry groups are projectively represented. The condensed-matter counterparts of the Poincar\'{e} group could be the spacetime groups of periodically driven crystals or spacetime crystals featuring spacetime periodicity. In this study, we establish the general theory of projective spacetime symmetry algebras of spacetime crystals, and reveal their intrinsic connections to gauge structures. As important applications, we exhaustively classify (1,1)D projective symmetry algebras and systematically construct spacetime lattice models for them all. Additionally, we present three consequences of projective spacetime symmetry that surpass ordinary theory: the electric Floquet-Bloch theorem, Kramers-like degeneracy of spinless Floquet crystals, and symmetry-enforced crossings in the Hamiltonian spectral flows. Our work provides both theoretical and experimental foundations to explore novel physics protected by projective spacetime symmetry of spacetime crystals.
\end{abstract}

\maketitle

\noindent \textbf{\large Introduction}\\
The Poincar\'{e} group as the symmetry group of relativistic spacetime is fundamental in high energy physics. In analyzing the representation of the Poincar\'{e} group by quantum matter, E. Wigner's revealed one of the most fundamental principles of quantum theory: symmetry groups are projectively represented in his seminal work~\cite{Wigner}. 

The condensed-matter analogues of the Poincar\'{e} group may be spacetime groups for various periodically driven crystals or spacetime crystals with finite spacetime periodicity~\cite{Floquet1,Floquet2,Floquet3,Floquet4,Floquet5,Floquet_ano,Floquet_table, Yaoshunyu, Floquet_indicator, Floquet_review2,Floquet_review1,Congjun}.  These groups are extensions of space groups with finite spacetime-translational symmetry, and are commonly used to analyze spacetime crystals~\cite{Congjun,cry_pointgroup,Vishwanath_glide, Niu, YangPeng}. However, the theory of projective representations of spacetime groups (PRSTGs) has yet to be established. Nevertheless, its fundamental importance is evident, as Wigner's philosophy suggests that every spacetime crystal, regardless of whether or not it has interactions, should projectively represent the spacetime group. 

On the other hand, the broad range of applications for the theory may be seen from the fact that PRSTGs can be induced from gauge fluxes over spacetime lattices. spacetime crystals are mainly experimented by artificial crystals, most of which feature engineerable gauge fields. Additionally, gauge structures can emerge from the periodic evolution of spacetime crystals~\cite{lattice_Floquet}. As a result, PRSTGs ubiquitously exist in spacetime crystals, and the theory of PRSTGs is expected to generate widespread interests, given the variety of artificial crystals available, including photonic and acoustic crystals, cold atoms in optical lattices, periodic mechanical systems and electric-circuit~\cite{Ozawa2019rmp,MaGuancong2019nrp,Lu2014,Yang2015,Acoustic_Crystal,Imhof2018,Yu2020,Prodan_Spring,Huber2016,Cooper2019,Optical_Lattice_RMP}. Notably, the projective representations of space groups for static crystals with gauge structures have recently attracted intensive attention in the field of topological physics ~\cite{switch,k.p,Klein,Mobius1,Mobius2,nonabelianflux,Hafstadter_Bernevig,proPT, Zhiyi}. 

%Since both spacetime symmetry and gauge structure are fundamentally significant for time crystals, here we present a general theory for projective spacetime symmetry of time crystals with gauge structures.

This work establishes the general theory of projective representations of PRSTGs and approaches it from several perspectives. The first step is to develop a method for classifying all projective representations of a given spacetime group. We then apply this method to exhaustively classify all $(1,1)$D spacetime groups, and characterize each class of projective representations using projective symmetry algebras (PSAs). The PSAs are particularly useful for practical applications. The classifications and corresponding PSAs are tabulated in Tab.~1. Based on these PSAs, we develop systematical methods to construct a canonical spacetime-lattice model for each $(1,1)$D spacetime group. By appropriately varying the flux configuration of the canonical model, it is possible to realize all PSAs of the spacetime group. It is noteworthy that recently the classification of PSAs for static crystals with time-reversal invariance has been made in Ref.~\cite{Zhiyi}. Besides different symmetry settings and physical systems, the classification of PSAs for spacetime crystals bears an important technical difference compared with the classification in  Ref. \cite{Zhiyi}, i.e., all spacetime symmetries reversing the time direction are anti-unitary.

Our research provides a fundamental framework for investigating the unique physics of spacetime crystals that may emerge from gauge-field enriched spacetime symmetry. Our findings include several noteworthy results, such as the electric Floquet-Bloch theorem, Kramers-like degeneracy of spinless spacetime crystals, and symmetry-enforced crossings in the Hamiltonian spectral flows. All these discoveries surpass the conventional theory of spacetime symmetry. Hence, upon the theoretical foundation established here, we expect  a multitude of novel physics resulting from projective spacetime symmetry to be explored by the community.

\bigskip
\noindent \textbf{\large Results}

\noindent{\textbf{Projective spacetime symmetry algebras}} \\
Let us start with recalling some basics of spacetime groups. Since the time translation operator ${L}_T$ is contained in any spacetime group, each $(d,1)$D spacetime group is isomorphic to a $(d+1)$D space group. However, not every $(d+1)$D space group can be interpreted as a $(d,1)$D spacetime group~\cite{Congjun}. This is because symmetry operations cannot mix spatial and temporal dimensions, i.e., time reversal, time-glide reflection and time-screw rotation operations are allowed, but spacetime rotations more than twofold are forbidden. Hence, only a subset of $(d+1)$D space groups can be interpreted as spacetime groups of $H(t)$. For $(1,1)$D time crystals, all possible spacetime groups are listed in Tab.~1, using notation adapted from $2$D space groups.

\begin{figure*}[ht]
	\centering
	\includegraphics[width=\textwidth]{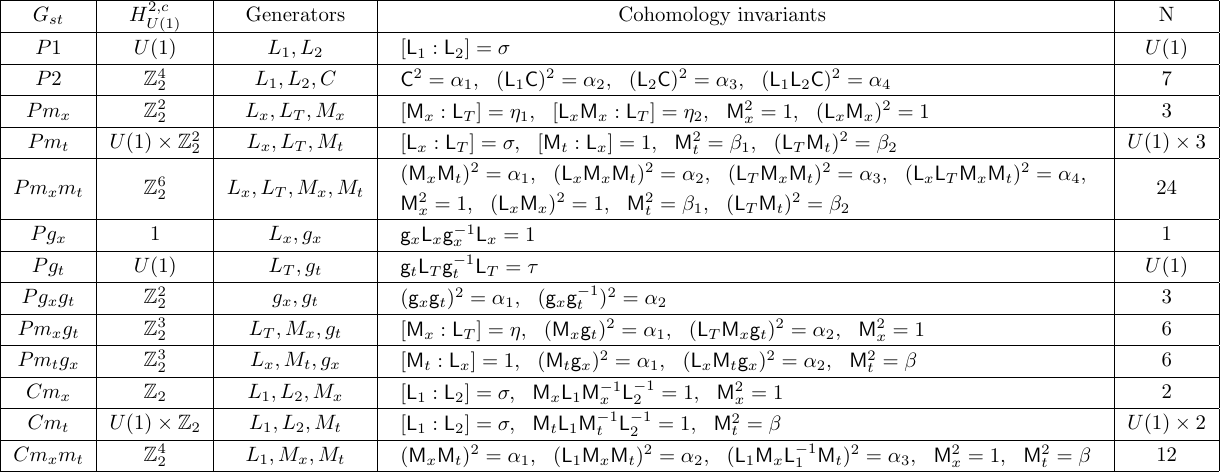}
	\captionsetup{justification=raggedright}
	\caption*{\textbf{Tab. 1. $U(1)$-projective algebras of 13 $(1,1)$D spacetime groups. } The columns in order list the underlying spacetime groups $G_{st}$, the second cohomology group $H^{2,c}(G_{st},U(1))$, the selected generators, the modified algebraic relations, and the numbers $N$ of isomorphic classes of algebras. The multiplicative commutator is defined as $[A:B]=ABA^{-1}B^{-1}$. $L_1$ and $L_2$ denote two primitive space-time translations. $M_x$ and $M_t$ are space and time reflections, respectively. $C=M_xM_t$ denotes the spacetime inversion. $g_x=L_{T/2}M_x$ and $g_t=L_{x/2}M_t$ denote space-glide reflection and time-glide reflection, respectively. Factors $\alpha, \beta, \eta$ are valued in $\mathbb{Z}_2=\{\pm 1\}$, and factors $\tau,\sigma$ are valued in $U(1)$ except the $\sigma$ factor for $Cm_x$, which is valued in $\mathbb{Z}_2$.}
	\label{Table}
\end{figure*}

Consider a spacetime group $G_{st}$. Under a projective representation,  the multiplication rules of the symmetry operators $\mathsf{S}$ are modified with additional phase factors. If $S_0S_1=S_{2}$, then $\mathsf{S}_0\mathsf{S}_1=\Omega(S_0,S_1)\mathsf{S}_{2}$, where $S_i$ are symmetry elements in $G_{st}$. $\Omega$ is called the multiplier of the projective representation and is valued $U(1)$. Linear and anti-linear operators always satisfy the associativity, i.e., $(\mathsf{S}_1 \mathsf{S}_2)\mathsf{S}_3=\mathsf{S}_1(\mathsf{S}_2\mathsf{S}_3)$ for three arbitrary operators $\mathsf{S}_1$, $\mathsf{S}_2$ and $\mathsf{S}_3$. Hence, the multiplier $\Omega$ satisfies the equation, 
\begin{equation}
\Omega(S_1,S_2)\Omega(S_1S_2,S_3)=\Omega(S_1,S_2S_3)c_{S^1}[\Omega(S_2,S_3)].
\end{equation}
Here, $c_S$ is an operator which depends on $S$. $c_S:=1$ if $S$ is unitary, $c_S:=\mathcal{K}$ if $S$ is antiunitary, where $\mathcal{K}$ is the complex conjugation operator. For spacetime groups, $S$ is antiunitary if it reverses time. Meanwhile, one may modify the phase of each $\mathsf{S}$ by $\chi(S)\mathsf{S}$ with $\chi(S)\in U(1)$. This transforms the multiplier to an equivalent one, i.e.,  
\begin{equation}
\Omega(S_1,S_2)\sim \Omega(S_1,S_2)~c_{S_1}[\chi(S_2)]\chi(S_1)/\chi(S_1S_2).
\end{equation}
The equivalence classes of multipliers form an Abelian group $H^{2,c}(G_{st},U(1))$, called the twisted second cohomology group of $G_{st}$~\cite{MooreAG}. It is clear that all projective representations of $G_{st}$ are classified by $H^{2,c}(G_{st},U(1))$. For $(1,1)$D spacetime crystals, we work out all the classifications as listed in the second column of Tab.~1.
Notably, because of the anti-unitarity of time reversal, the $(d,1)$D spacetime groups have distinct projective representations compared with the corresponding $(d+1)$D space groups. 

Under projective representations, the original group algebraic relations of symmetry operators are modified by $\Omega$, resulting in the modified algebras called the PSAs~\cite{switch,k.p,Klein,Mobius1,Mobius2,Hafstadter_Bernevig}. These algebras are classified by $H^{2,c}(G_{st},U(1))$, which gives us the number of algebraic classes. However, for physical applications, it is essential to exhibit the explicit forms of these algebras. To achieve this in $(1,1)$ dimensions, we carefully select a set of generators (in the third column of Tab.~1) for each $G_{st}$, such that each PSA corresponds to a modification of the corresponding multiplication relations by a set of independent phase factors (in the fourth column of Tab.~1). The derivation for all these PSAs can be found in the Supplementary Information (SI).

For example, $Pm_x$ can be generated by unit translations $L_x, L_T$ and spatial reflection $M_x$, which satisfy the following relations: $[M_x : L_T]=1$, $[L_xM_x : L_T]=1, M_x^2=1$ and $(L_xM_x)^2=1$. All the PSAs of $Pm_x$ can be obtained by modifying  these relations with $U(1)$ phase factors. Specifically, $[\mathsf{M}_x : \mathsf{L}_T]=\eta_1$, $[\mathsf{L}_x\mathsf{M}_x : \mathsf{L}_T]=\eta_2$, $\mathsf{M}_x^2=\gamma_1$ and $(\mathsf{L}_x\mathsf{M}_x)^2=\gamma_2$. However, phase factors $\gamma_1,\gamma_2$ can be set to be 1 by redefining $\mathsf{L}_x \to \mathsf{L}'_x=\gamma_1^{1/2}\gamma_2^{-1/2}  \mathsf{L}_x$, $\mathsf{M}_x \to \mathsf{M}'_x =\gamma_1^{-1/2}  \mathsf{M}_x$. Moreover, phase factors $\eta_1$, $\eta_2$ can only take values in $\mathbb{Z}_2=\{\pm 1\}$, since $\mathsf{L}_T \gamma_1 =  \mathsf{L}_T \mathsf{M}_x^2 =  \eta_1^{-2} \mathsf{M}_x^2 \mathsf{L}_T=\eta_1^{-2}\mathsf{L}_T\gamma_1$ and $\mathsf{L}_T \gamma_2 =  \mathsf{L}_T (\mathsf{L}_x\mathsf{M}_x)^2  =  \eta_2^{-2}(\mathsf{L}_x \mathsf{M}_x)^2\mathsf{L}_T=\eta_2^{-2}\mathsf{L}_T \gamma_2$, leading to $\eta_1^2=\eta_2^2=1$. All the possible values of phase factors, i.e., $\eta_1$, $\eta_2=\pm 1$ correspond to the $\mathbb{Z}_2^2$ classification of the projective symmetry algebras of $Pm_x$. 

It is significant to note that all phase factors in Tab.~1 are invariant under multiplying each symmetry operator by an arbitrary phase factor $\chi$. Therefore, they are solely determined by the equivalence class. Conversely, for each $G_{st}$, the free phase factors in Tab.~1 are complete, i.e., sufficient to specify equivalence classes of $H^{2,c}(G_{st},U(1))$. Thus, for each $G_{st}$ we call these phase factors a complete set of cohomology invariants.

\begin{figure*}[t]
	\centering
	\includegraphics[width=0.8\textwidth]{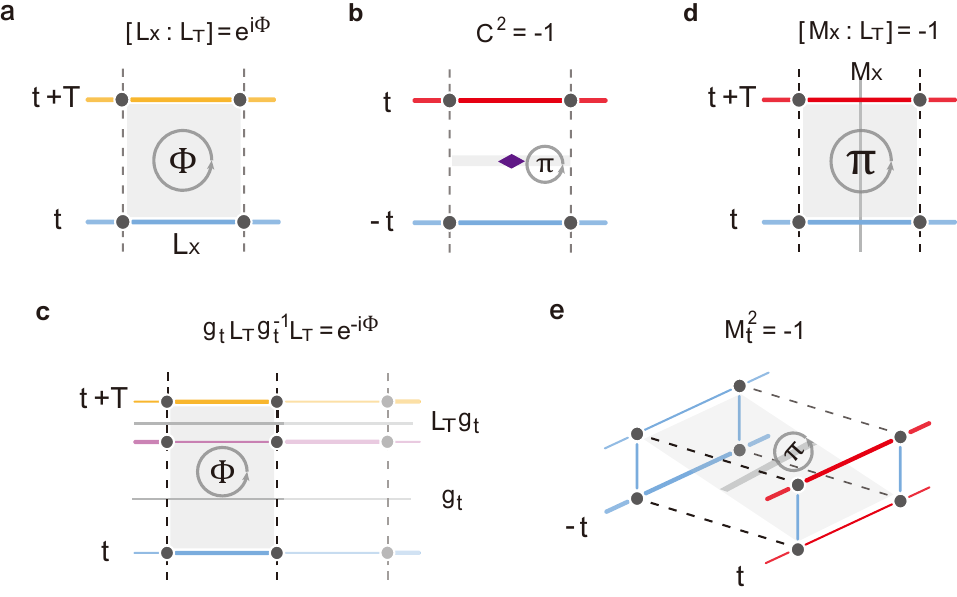}
	\captionsetup{justification=raggedright}
\caption*{\textbf{Fig. 1. Realization of projective algebraic relations. } \textbf{a,b,c,d,e} illustrate the realization of nontrivial cohomology invariants $\sigma, \alpha, \eta, \tau, \beta$ respectively. The dashed lines indicate the evolution of hopping amplitudes with time. The dark areas indicate fluxes, which are realized by tunning hopping amplitudes.
The red hoppings differ the corresponding blue hoppings by a minus sign. The yellow and purple hoppings differ the corresponding blue hoppings by a phase. }
	\label{fig1}
\end{figure*}

Nevertheless, two elements in $H^{2,c}(G_{st},U(1))$ may correspond to isomorphic algebras. The number of independent algebras for each spacetime group is given in the last column of Tab.~1. Let $Pm_x$ be an example. $H^{2,c}(Pm_x,U(1))=\mathbb{Z}_2^2$ consists of four possibilities specified by $[\mathsf{M}_x:\mathsf{L}_T]=\eta_1=\pm 1$ and $[\mathsf{L}_x\mathsf{M}_x:\mathsf{L}_T]=\eta_2=\pm 1$. $\mathsf{M}_x$ and $\mathsf{L}_x\mathsf{M}_x$, as reflections through parallel lines related by half a translation, are on an equal footing. Hence, $(\eta_1,\eta_2)=(-1,1)$ and $(\eta_1,\eta_2)=(1,-1)$ correspond to isomorphic algebras, leading to $3$ isomorphic classes.\\

\noindent{\textbf{Realization by spacetime gauge fluxes}} \\
We proceed to construct spacetime crystal models that can realize the PSAs by gauge fluxes. The general form of the time-periodic Hamiltonian is given by $H(t)=\sum_{ij} h_{ij}(t)e^{iA_{ij}(t)}c^\dagger_{i} c_j+\mathrm{H.c.}$, which can be succinctly written as $H(t)=\sum_{ij} c_i^\dagger H_{ij}(t) c_j$. Here, the subscripts $i,j$ indicate the positions of lattice sites in space, upon which the spatial symmetries act.

Symmetry group consists of transformations that leave the system invariant. With  gauge degrees of freedom, a symmetry transformation may map the system to a gauge equivalent one, which in general is different from the original system. An elementary example is that under the time translation $L_T$, the Hamiltonian $H(t)$ is mapped to $H(t+T)$, and $H(t)$ and $H(t+T)$ are related by a gauge transformation $\mathsf{G}_T$, i.e.,
\begin{equation}\label{eq:periodicity}
	H(t+T)=\mathsf{G}_T^{-1} H(t) \mathsf{G}_T.
\end{equation}
This resembles the spacetime cylinder frequently considered for quantum gauge anomalies~\cite{WittenSU(2)}, and moreover, we shall see analogous crossings in the spectral-flow diagram later. Consequently, we can define the physical time-translation operator as $\mathsf{L}_T=\mathsf{G}_TL_T$ with $[H(t),\mathsf{L}_T]=0$. Similarly, for any symmetry $S\in G_{st}$, the physical symmetry operator is the combination $\mathsf{S}=\mathsf{G}_S S$ with $\mathsf{G}_S$ the corresponding gauge transformation, and it is  $\mathsf{S}$ that commutes with the Hamiltonian, i.e.,
\begin{equation}
	[H(t),\mathsf{S}]=0.
\end{equation}
Note that a gauge transformation is a diagonal matrix indexed by the lattice sites. Because of gauge transformations, the multiplication of two arbitrary symmetry operators $\mathsf{S}_1$ and $\mathsf{S}_2$ is modified by the multiplier $\Omega(S_1,S_2)=\mathsf{G}_{S_1}S_1\mathsf{G}_{S_2}S_1^{-1}\mathsf{G}_{S_1S_2}^{-1}$, which is a $U(1)$-factor as proved in the SI. It is noteworthy that the equivalent class of $\Omega(S_1,S_2)$ is completely determined by the flux configuration, i.e., two gauge equivalent gauge connection configurations correspond to equivalent multipliers.

We realize all PSAs for the 13 spacetime groups in Tab.~1 for both $U(1)$ and $\mathbb{Z}_2$ gauge fields. In Tab.~1, we have categorized the phase factors into five classes, denoted by $\sigma$, $\alpha$, $\beta$, $\eta$ and $\tau$. Each of these classes can be realized by lattice building blocks with appropriate gauge fluxes, which are illustrated in Fig.~1 and elucidated below. The technical details are provided in the SI.

$(i)$ The first class is $\sigma=[\mathsf{L}_1:\mathsf{L}_2]$ for the translation subgroup, where $\mathsf{L}_1$ and $\mathsf{L}_2$ are primitive spacetime translations. The value of $\sigma$ corresponds to the flux in a space-time loop formed by $\mathsf{L}_1\mathsf{L}_2\mathsf{L}_1^{-1}\mathsf{L}_2^{-1}$. Figure~1a shows the case when $\mathsf{L}_1=\mathsf{L}_x, \mathsf{L}_2=\mathsf{L}_T$. The general cases of spacetime translations can be found in the SI. 

$(ii)$ The second class concerns cohomology invariants for spacetime inversion, e.g., $\alpha=(\mathsf{M}_x\mathsf{M}_t)^2$. $\alpha=\pm 1$ corresponds to flux 0 or $\pi$ in every spacetime inversion invariant loop, as illustrated in Fig.~1b.

$(iii)$ The third class is denoted by $\tau$, relating time translations and time glide reflections, e.g., $\tau=\mathsf{g}_t\mathsf{L}_T\mathsf{g}_t^{-1}\mathsf{L}_T^{-1}$. The phase factor $\tau^*$ corresponds to the flux in a loop formed by $\mathsf{g}_t\mathsf{L}_T\mathsf{g}_t^{-1}\mathsf{L}_T^{-1}$, as illustrated in Fig.~1c.

$(iv)$ The fourth class denoted $\eta$ modifies the relation between spatial reflections and time translations, e.g., $\eta=\mathsf{M}_x\mathsf{L}_T\mathsf{M}_x^{-1}\mathsf{L}_T^{-1}$. $\eta=\pm 1$ corresponds to $0$ or $\pi$ flux in the loops formed by $\mathsf{M}_x\mathsf{L}_T\mathsf{M}_x^{-1}\mathsf{L}_T^{-1}$, as illustrated in Fig.~1d.

$(v)$ The fifth class corresponds to the square of a time reversal, i.e, $\beta=\mathsf{M}_t^2$ with $\beta=\pm 1$. $\beta=-1$ cannot be realized by one-layer model, here we realize it by a two layer-model, where the time reversal is effectively realized by the combination of time reversal and a horizontal two-fold rotation, as illustrated in Fig.~1e.

These building blocks  form the basis for systematic model constructions for all projective symmetry algebras in Tab.~1. Here we present the models for $P1$ and $Pg_t$, respectively, each of which has only one phase factor. The remaining models are available in the SI.

\textit{Projective P1 symmetry.} 
Our first example is simply $P1$ consisting of only spacetime translations. For simplicity, we choose symmetry operators as spatial and time translators, $L_x$ and $L_T$. We construct the lattice model with only the nearest-neighbor hopping amplitudes as shown in Fig. 2a. The momentum-space Hamiltonian is given by
\begin{equation}\label{Hp1}
	\begin{aligned}
		H(k,t)=\left[\begin{array}{ccc} 0& w_1(t)  & J(t) e^{-ika}\\ w_1(t) & 0 & w_2(t)\\ J(t) e^{ika} &w_2(t)& 0 \end{array}\right],
	\end{aligned}
\end{equation}
where $w_{1,2}$ are the two hopping amplitudes within each cell, and $J(t)$ is the hopping amplitude between two neighboring cells.

For simplicity, we choose $[\mathsf{L}_x : \mathsf{L}_T]=-1$, namely $\{\mathsf{L}_x, \mathsf{L}_T\}=0$, which can be realized by inserting odd times of $\pi$-flux in each spacetime plaquette formed by $L_x$ and $L_T$. This can be implemented by setting $w_{1,2}(t)=-w_{1,2}(t+T)$ and $J(t)=-J(t+T)$. In this setting, the flux in each spacetime unit cell is $3\pi$, and we note that any odd times of $\pi$-flux is appropriate and leads to the same projective algebra. Since the gauge condition preserves $L_x$, $\mathsf{L}_x=L_x$. But translating $H(k,t)$ by $L_T$ should be combined with a gauge transformation $\mathsf{G}_T$ (see Fig. 2a), which in momentum space is represented by
\begin{equation}\label{GT}
	{\mathsf{G}}_T=  M\mathcal{L}_{\frac{G}{2}},\quad  M=\left[\begin{array}{ccc} -1& 0 & 0\\ 0 & 1 & 0\\ 0 &0 & -1 \end{array}\right],
\end{equation}
where $\mathcal{L}_{\frac{G}{2}}$ denotes a half reciprocal translation, namely, sending $k$ to $k+G/2$. Then, it is easy to verify $\{\mathsf{L}_x, \mathsf{L}_T\}=0$ with ${\mathsf{L}}_T={\mathsf{G}}_T L_T$ and ${\mathsf{L}}_x=e^{ika}$.  The anti-commutativity can also be pictorially observed: In Fig.~2a, the signs $\pm$ of $\mathsf{G}_T$ are exchanged by $L_x$, i.e., $L_x\mathsf{G}_TL_x^{-1}=-\mathsf{G}_T$.

\textit{Projective $Pg_t$ symmetry.} Our second example is $Pg_t$ with two generators $g_t$ and $L_T$. In particular, $g_t$ is the glide time-reflection, which combines the time inversion $M_t$ with a half translation $L_{x/2}$ along the spatial lattice, namely $g_t=L_{x/2}M_t$. From Tab.~1, we consider the projective algebra, $\mathsf{g}_t \mathsf{L}_T \mathsf{g}_t^{-1} \mathsf{L}_T=-1$.

The model is constructed as in Fig.~2b, and is described by the Hamiltonian,
\begin{equation}\label{HP2}
	\begin{aligned}
		H(k,t)=\left[\begin{array}{cccc} 0& w_1(t) & 0 & J(t) e^{-ika}\\ w_1(t)& 0 & w_2(t)& 0\\ 0 &w_2(t)& 0 & w_3(t)\\ J(t)e^{ika} &0 &  w_3(t) &0\end{array}\right].
	\end{aligned}
\end{equation}
The nontrivial projective algebra can be realized by inserting a $\pi$-flux into each loop formed by $g_tL_Tg_t^{-1}L_T$ (see Fig.~2b). Particularly, in Eq.~\eqref{HP2}, this is satisfied by $w_1(t)=-w_3(-t), w_2(t)=J(-t)$ and $w_1(t)=-w_1(t+T),\ w_2(t)=w_2(t+T),\ w_3(t)=-w_3(t+T)$ and $J(t)=J(t+T)$. Now, $g_t$ should be accompanied by the gauge transformation $\mathsf{G}_{g_t}$ (see Fig.~2b). In momentum space, ${\mathsf{G}}_{g_t}=\sigma_3 \otimes \sigma_3$ with $\sigma$'s the Pauli matrices, and
\begin{equation} \label{gt}
	{\mathsf{g}}_t={\mathsf{G}}_{g_t}{g}_t=M_g \mathcal{I}_k\mathcal{I}_t \mathcal{K}, ~ M_g(k)=\left[\begin{array}{cc} 0& \sigma_3\\ -e^{ika}\sigma_3 & 0  \end{array}\right],
\end{equation}
where $\mathcal{I}_{k}$ ($\mathcal{I}_{t}$) is the inversion of $k$ ($t$), and $\mathcal{K}$ the complex conjugation. Moreover, one can check that the gauge transformation $\mathsf{G}_T$ for $L_T$ is equal to $\mathsf{G}_{g_t}$, i.e., ${\mathsf{L}}_T=\sigma_3\otimes \sigma_3 L_T$. It is straightforward to verify that ${\mathsf{g}}_t H(k,t){\mathsf{g}}_t^{-1}={\mathsf{L}}_T H(k,t){\mathsf{L}}_T^{-1}=H(k,t)$, and the projective algebra ${\mathsf{g}}_t{\mathsf{L}}_T {\mathsf{g}}_t^{-1}{\mathsf{L}}_T=-1$.\\

\begin{figure*}[ht]
	\centering
	\includegraphics[width=0.9\textwidth]{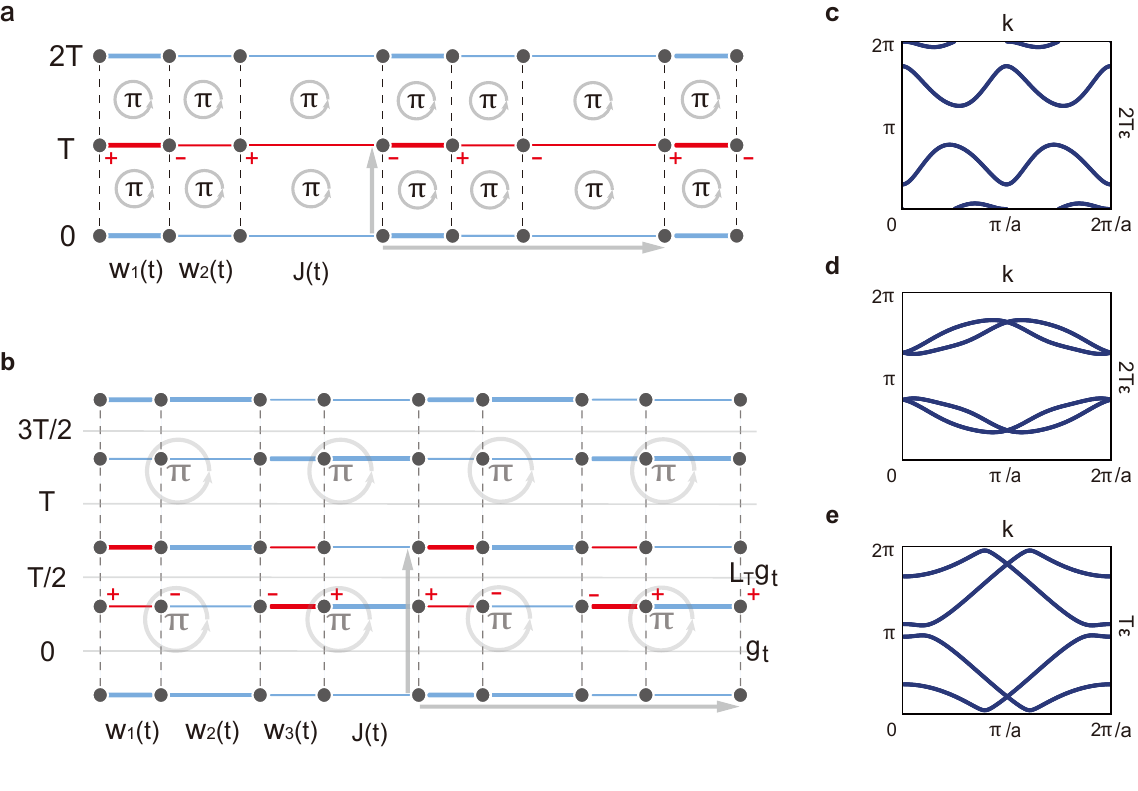}
	\captionsetup{justification=raggedright} 
	\caption*{\textbf{Fig. 2  Projective $P1$ and $Pg_t$ symmetry model.} \textbf{a} Spacetime tight-binding model with projective algebra of $P1$.  The dashed lines indicate a continuous evolution of hopping amplitudes. The three hopping amplitudes are explicitly shown at $t=0,T, 2T$. %Every hopping amplitude acquires a minus sign after a period $T$, and therefore each spacetime plaquette has flux $\pi$. 
The signs marked in red at each site at $t=T$ describe the gauge transformation $\mathsf{G}_T$ needed to restore the initial connections. \textbf{b} Spacetime tight-binding model with projective representation of $Pg_t$. Glide axes are plotted as dark horizontal lines. The distribution of hopping amplitudes is explicitly marked at five times, respectively. We  observe that  the glide time-reflection through $t=T/2$ is manifestly preserved, while that for $t=0$ or $T$ is preserved up to a gauge transformation $\mathsf{G}_{g_t}$. The signs for $\mathsf{G}_{g_t}$ at a given time are marked in red.  The arrows in \textbf{a} and \textbf{b} are the primitive lattice vectors correspond to $L_x$ and $L_T$. \textbf{c} Quasibands of $U(2T)$ of the model.  \textbf{d} Quasibands of $U(2T)$ for the model in (\textbf{b}). \textbf{e} Quasibands of Floquet model with ordinary $Pg_t$ symmetry. The parameter values for (\textbf{c}), (\textbf{d}) and (\textbf{e}) are given in the SI. }
	\label{fig2}
\end{figure*}

\noindent{\textbf{Physical consequences of projective spacetime symmetry}} \\
We proceed to discuss the aforementioned three fascinating consequences of the projective spacetime symmetry algebras.

 \textit{(1) Electric Floquet-Bloch theorem.}  The electric Floquet-Bloch theorem for projective spacetime translational symmetry is the counterpart of the magnetic Bloch theorem for static systems with projective spatial translational symmetry in uniform magnetic fields~\cite{Floquet_review1, magneticBloch1,magneticBloch2}.  Without loss of generality, let us consider a $(1,1)$D spacetime crystal with flux $2\pi p/q$ through each spacetime unit cell. Then, the proper time and lattice translation operators $\mathsf{L}_T$ and $\mathsf{L}_x$ that commute with the Hamiltonian satisfy the gauge-invariant commutation relation $[\mathsf{L}_x: \mathsf{L}_T]=e^{i2\pi p/q}$. We choose two commuting operators $\mathsf{L}_x$ and $(\mathsf{L}_T)^q$, which have the common eigenstates corresponding to the quantum numbers $(k, \epsilon_n)$. Note that $\epsilon_n$ denote the eigenvalues of $U(k,qT)=\mathcal{T}\exp[ -i\int_0^{qT}H(k,t) dt]$, which are known as ``quasi-energies'', with $\mathcal{T}$ indicating the time ordering. The Floquet-Bloch states labeled by $(k, \epsilon_n)$ can be written as
\begin{equation}
	\psi_{k,n}(x,t)=e^{i(kx-\epsilon_{n}(k) t)}u_{k,n}(x,t),
\end{equation}
where $u_{k,n}$ satisfies
\begin{equation*}
u_{k,n}(x+a, t)=u_{k,n}(x,t),~~ u_{k,n}(x,t+qT)=\lambda(x)u_{k,n}(x,t).
\end{equation*}
Here, the phase $\lambda(x)=e^{-2\pi i px/a}$ arises from the assumption of a uniform electric field, and vanishes at each lattice site. The proof of the electric Floquet-Bloch theorem is given in the Methods.

It is noteworthy that $\psi_{k,n}$, $\mathsf{L}_T\psi_{k,n}, ...$, $(\mathsf{L}_T)^{q-1} \psi_{k,n}$ are $q$-fold degenerate states for the quasi-energy, where $(\mathsf{L}_T)^m  \psi_{k,n}$ has momentum $k+2\pi mp/qa$. This is exemplified by the model \eqref{Hp1}. If $\psi_{k,n}$ is an eigenstate of $U(k,2T)$, then $U^{-1}(k+\pi/a,T)M^{-1}\psi_{k,n}$ is an eigenstate of $U(k+\pi/a, 2T)$ with the same eigenvalue. Accordingly, as observed in Fig.~2c, the quasi-energy spectrum is the same at $k$ and $k+\pi/a$, and hence has a half period.

\textit{(2) Kramers degeneracy protected by projective symmetry.} For static systems, it has been shown that projective symmetries can lead to the Kramers degeneracy for spinless systems~\cite{switch}. Here, we show that this extraordinary phenomenon also occurs in the quasi-energy spectra for time crystals with projective symmetry algebras.

Let us illustrate this by $Pg_t$ symmetry of \eqref{HP2}, while another example for group $P2$ can be found in the SI.  The symmetry operator  \eqref{gt} constrains the Hamiltonian as $M_g H(k,t) M_g^{-1}=H^*(-k,-t)$, which leads to the evolution operator $U(k,2T)=\mathcal{T}\exp[-i\int_0^{2T} dt H(k,t)]$ satisfying
\begin{equation}
	M_g U(k,2T)M_g^{-1}=U^{-1*}(-k,2T).
\end{equation}
Hence, for $U(k,2T)$, we should consider the anti-unitary symmetry operator $M_g(k)\mathcal{K}$ with $(M_g(k)\mathcal{K})^2=-e^{ika}$. It is significant to notice that the square of $M_g(k)\mathcal{K}$ equals $-1$ at $k=0$, which leads to the Kramers degeneracy at $k=0$ for the quasi-energy.

Likewise, one can show that another time glide symmetry $\mathsf{L}_T\mathsf{g}_t$ protects the Kramers degeneracy at $k=\pi/a$. The Kramers degeneracies at both $k=0$ and $\pi/a$ can be seen in Fig.~2d. It is noteworthy that the Kramers degeneracy at $k=\pi/a$ also appears for ordinary $Pg_t$ group ~\cite{Congjun} (see Fig.~2e), but that at $k=0$ only appears for projective $Pg_t$ algebra.

\textit{(3) Symmetry-enforced crossings of the Hamiltonian spectral flow.} We first note that a sublattice symmetry $\Gamma$ operates on a bipartite lattice exactly as a $\mathbb{Z}_2$ gauge transformation: Each A-sublattice (B-sublattice) site is multiplied by a sign $+$ ($-$). Let us consider a periodic time evolution restores the crystalline system up to the gauge transformation $\Gamma$ specified by the sublattice symmetry, i.e., $H(k,T)=\Gamma H(k,0)\Gamma$. Since $\{\Gamma, H\}=0$ for sublattice symmetry, we have $H(k,T)=-H(k,0)$. Thus, generally, for a gapped $H(k,0)$, a continuous $H(k,t)$ with respect to $t$ exchanges valence and conduction bands in a period $T$, which enforces energy band crossings intermediately. Note that here we consider the instantaneous spectrum of the Hamiltonian $H(t)$ rather than the quasienergy spectrum. Above band crossing is illustrated by a $t$-dependent dimerized model as a generalization of the Su-Schrieffer-Heeger model in the Methods.\\

\noindent\textbf{\large Discussion}\\
To summarize, our work presents a general theory for the projective spacetime symmetry of spacetime crystals. We have achieved a complete classification and presented the PSAs of $(1,1)$D spacetime crystals, and all systematically constructed models can be easily realized through the use of engineerable gauge fields in artificial crystals. Although our focus has been on spinless systems, the extension of our theory to spinful systems is a straightforward matter. In such cases, spin-1/2 will contribute a phase of -1 to the squares of spacetime rotations and time reversals. Therefore, to include spin degrees of freedom, we can simply replace $\alpha$ and $\beta$ in Tab. 1 with $\alpha=(-1)^{2s}\alpha$ and $\beta=(-1)^{2s}\beta$, respectively.

%\textcolor{blue}{For simplicity, the spacetime lattice models we demonstrated all have primitive lattice vectors $L_x$ and $L_T$. However, for some spacetime groups like $P1$, primitive lattice vectors can be space-time mixed, which can lead to some physical consequences \cite{YangPeng, Niu}. Our framework of projective symmetry is still appliable to these models and the physical consequences remaine to be explored. }

The present work provides a comprehensive framework for investigating the fascinating physics of spacetime crystals with projective spacetime symmetry. Through a thorough analysis of the PSAs, we have uncovered three significant physical consequences. Moving forward, an intriguing direction for future research would be to explore spacetime topological phases with projective symmetry. This research avenue is particularly compelling given the systematic classifications of crystalline topological phases that have emerged from the study of static crystal symmetries~\cite{Fu_crystal_topo,Sato,symmetry_indicator2,Fang_nature,high-quality_nature,Wan_nature}.

To realize spacetime lattices models with projective symmetries, one need to realize the required time-dependent hopping amplitudes, which may be done by cold atoms, photonic crystals, ultrafast spintronics or other systems \cite{ex1,ex2,ex3,ex4,ex5,ex6}.

Finally, it is worth noting that while gauge fluxes on spacetime lattices can be used to realize PSAs, they may not be capable of representing every possible PSA in higher dimensions. Certain PSAs may necessitate intrinsic many-body states, indicating a potential avenue for further exploration.

%\newpage
%~\newpage
\bigskip
\noindent \textbf{\large Methods}\\
\noindent\textbf{Proof of the electric Floquet-Bloch theorem}\\
Here, we prove the Floquet-Bloch theorem in a uniform electric field, which we call electric Floquet-Bloch theorem. We only concern the 1+1D case, while the generazation to 3+1D is straightforward.  

The Hamiltonian for a periodic driving system in a uniform electric field $E_x=E$ can be written as
\begin{equation}
H(x,t)=-\frac{1}{2m} (\partial_x-iA_x(t))^2 + U(x,t).
\end{equation}
where the potential $U(x,t)$ is periodic, i.e, $U(x+a,t)=U(x,t+T)=U(x,t)$ and we choose the  gauge $A_0=0,\ A_x(t)=-Et$. Due to the electric field, 
$H(x,t)$ is not periodic at the time direction, i.e., $H(t+T)\not = H(t)$. If we define two translation operators $L_x, L_T$ by $L_x f(x,t)=f(x+a,t)$ and $L_T f(x,t)=f(x,t+T)$, then $H(x,t)$ commutes with $L_x$ but not with $L_T$. However, we observe that $H(x,t+T)$ only differs from $H(x,t)$ by a gauge transformation $A_x \to A_x +\partial_x \chi(x), \ \chi(x)=ETx$. This gauge transformation can be equivalently defined as 
\begin{equation}
H(x,t) \to G_T H(x,t)G_T^{-1}=e^{i\chi(x)}H(x,t)e^{-i\chi(x)},
\end{equation} 
where $G_T=e^{i\chi(x)}$. So we can define a proper time translation operator $\mathsf{L}_T=G_TL_T$, which commutes with $H(x,t)$, $\mathsf{L}_T H(x,t)\mathsf{L}_T^{-1}=H(x,t)$. One can check the two proper translation operators $\mathsf{L}_T$ and $L_x$ satisfy
\begin{equation}
L_x\mathsf{L}_T{L}_x^{-1}\mathsf{L}_T^{-1}=e^{iEaT}=e^{i\Phi_E}.
\end{equation}
We assume the electric flux in every unit cell is  $\Phi_E=p\Phi_0/q=2\pi p/q$. 

To find the constrain of the translation symmetry on the wavefunction, we have to find commutative operators which commute with the Hamiltonian. Here, we can take $L_x, (\mathsf{L}_T)^q$ as two generators, they generate an abelian spacetime translation group $G$. We define
\begin{equation}
\mathcal{H}(x,t)=H(x,t)-i\partial_t,
\end{equation}
then solutions satisfy $\mathcal{H}(x,t)\psi(x,t)=0$. All solutions form a solution space $V$. Because every group element $g\in G$ commutes with $\mathcal{H}(x,t)$, the solution space $V$ is a representation of $G$, which can be decomposed by irreducible representations of $G$. Since $G$ is abelian, its irreducible representations are one dimensional, which are labelled by $(k,\epsilon)$, with character $\chi_{k,\epsilon}((L_x)^n (\mathsf{L}_T^q)^{m})=e^{ikna-i\epsilon qmT}$, $m,n \in \mathbb{Z}$.  The range of $(k,\epsilon)$ is $\ k\in [0,2\pi/a),\ \epsilon \in [0,2\pi/qT)$. If a solution $\psi(x,t)$ is in an irreducible representation labelled by $(k,\epsilon)$, it satisfies
\begin{equation}
\begin{aligned}
&L_x \psi(x,t)=e^{ika}\psi(x,t),\\
&(\mathsf{L}_T)^q \psi(x,t)=e^{-i\epsilon qT}\psi(x,t),
\end{aligned}
\end{equation}
that is,
\begin{equation}
\begin{aligned}
&\psi(x+a,t)=e^{ika}\psi(x,t),\\
& \psi(x,t+qT)=e^{-i2\pi p x/a}e^{-i\epsilon qT}\psi(x,t).
\end{aligned}
\end{equation}
We can rewrite it as
\begin{equation}
\begin{aligned}
&\psi(x,t)=e^{ikx-i \epsilon t} u(x,t),\\
&u(x+a,t)=u(x,t),\\
&u(x,t+qT)=e^{-i2\pi px/a}u(x,t).
\end{aligned}
\end{equation}
Plug this ansatz into  $\mathcal{H}(x,t)\psi(x,t)=0$, we can obtain a set of eigenvalues $\epsilon_{n}(k)$ and eigenstates $u_{n,k}(x,t)$. Thus we can label a state with by $(k,n)$ and we complete the proof of the electric Floquet-Bloch theorem.

Since $[\mathcal{H}(x,t), \mathsf{L}_T]$=0, if $\psi_{k,n}(x,t)$ is a solution, $\psi'(x,t)=\mathsf{L}_T\psi_{k,n}(x,t)=e^{i\chi(x)}\psi_{k,n}(x,t+T)$ is also a solution. Moreover, $\psi'(x,t)$ also has the quasienergy $\epsilon_n(k)$, but has momentum $k+2\pi p/qa$, this can be seen by
\begin{equation}
\begin{aligned}
(\mathsf{L}_T)^q \psi'(x,t)&=\mathsf{L}_T(\mathsf{L}_T)^q\psi_{k,n}(x,t)=e^{-i\epsilon_n(k)qT}\psi'(x,t),\\
L_x\psi'(x,t)&=L_x\mathsf{L}_T\psi_{k,n}(x,t)=e^{i\Phi_E}\mathsf{L}_TL_x\psi_{k,n}(x,t)\\&=e^{i(k+2\pi p/qa)a} \psi'(x,t).
\end{aligned}
\end{equation}
With this reason, $\mathsf{L}_T^m \psi_{k,n}(x,t), \ m=0,1,2,...,q-1$ are $q$-fold degenerate Floquet-Bloch states, wherein $\mathsf{L}_T^m \psi_{k,n}(x,t)$ has momentum $k+2\pi mp/qa$.\\

\noindent{\textbf{Band crossing due to projective symmetry}}\\
Here, we consider a driven  Su-Schrieffer-Heeger (SSH) model
\begin{equation}
H(t)=v(t)\sum_{i} c_{i,A}^\dagger c_{i,B}+ w(t)\sum_{i} c_{i,B}^\dagger c_{i+1,A} + h.c.
\end{equation}
where $v(t)$ and $w(t)$ are real. We require $H(t)$ to evolve adiabatically. If this model has projective time translation symmetry $H(t+T)=-H(t)$, (which can be written as $GH(t+T)G^{-1}=H(t)$, where $G=\mathrm{diag}(+1,-1,+1,-1,...)$ in real space), the instantaneous bands must cross at some time $t$. This is easy to see: In one period $T$, since $w(t+T)=-w(t), v(t+T)=-v(t)$, there must be $w(t_*)=v(t_*)$ at some  $t_*$. And we know that SSH model is gapless when $|w|=|v|$, so there is band crossing at $t_*$ for this driven SSH model.

\bigskip
\noindent \textbf{\large Data availability}\\
The data generated and analyzed during this study are available from the corresponding author upon request.

\bigskip
\noindent \textbf{\large Code availability}\\
All code used to generate the plotted band structures is available from the corresponding author upon request.

\bigskip
\def\bibsection{\ }
\noindent \textbf{\large References}
\bibliographystyle{naturemag}

%\bibliography{T_Crystal_ref}

\bigskip

\noindent \textbf{\large Acknowledgements}\\
This work is supported by National
Natural Science Foundation of China (Grants
No. 12161160315 and No. 12174181), and Basic Research
Program of Jiangsu Province (Grant No. BK20211506). 

\bigskip
\noindent \textbf{\large Author contributions}\\
Y.Z. conceived the idea and supervised the project. Z.Z. and Z.C. did the theoretical
analysis. Z.Z. and Y.Z. wrote the manuscript.

\bigskip
\noindent \textbf{\large Competing interests}\\
The authors declare no competing interests.

\newpage

\AtEndDocument{\includepdf[pages=1]{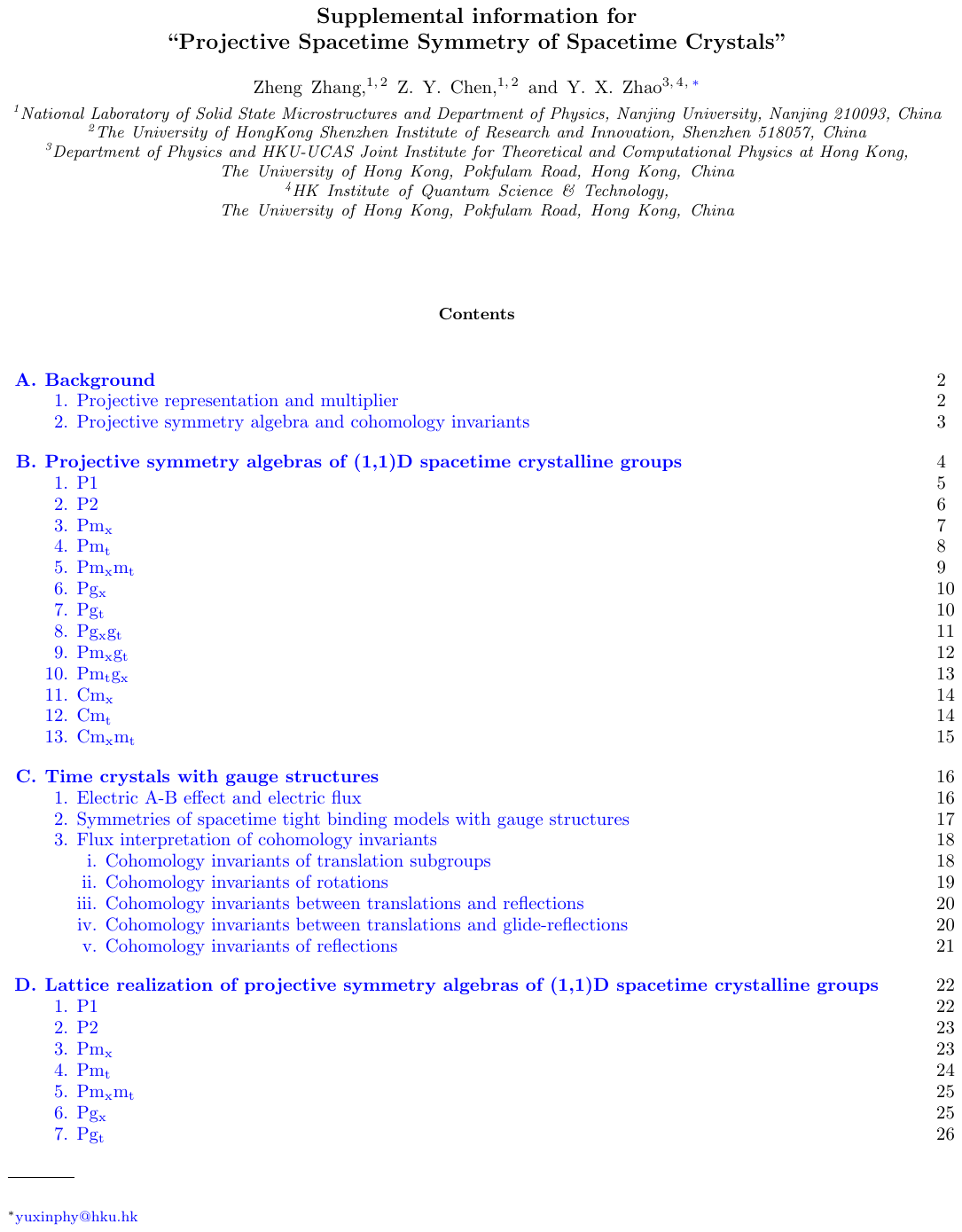}}
\AtEndDocument{\includepdf[pages=2]{supp.pdf}}
\AtEndDocument{\includepdf[pages=3]{supp.pdf}}
\AtEndDocument{\includepdf[pages=4]{supp.pdf}}
\AtEndDocument{\includepdf[pages=5]{supp.pdf}}
\AtEndDocument{\includepdf[pages=6]{supp.pdf}}
\AtEndDocument{\includepdf[pages=7]{supp.pdf}}
\AtEndDocument{\includepdf[pages=8]{supp.pdf}}
\AtEndDocument{\includepdf[pages=9]{supp.pdf}}
\AtEndDocument{\includepdf[pages=10]{supp.pdf}}
\AtEndDocument{\includepdf[pages=11]{supp.pdf}}
\AtEndDocument{\includepdf[pages=12]{supp.pdf}}
\AtEndDocument{\includepdf[pages=13]{supp.pdf}}
\AtEndDocument{\includepdf[pages=14]{supp.pdf}}
\AtEndDocument{\includepdf[pages=15]{supp.pdf}}
\AtEndDocument{\includepdf[pages=16]{supp.pdf}}
\AtEndDocument{\includepdf[pages=17]{supp.pdf}}
\AtEndDocument{\includepdf[pages=18]{supp.pdf}}
\AtEndDocument{\includepdf[pages=19]{supp.pdf}}
\AtEndDocument{\includepdf[pages=20]{supp.pdf}}
\AtEndDocument{\includepdf[pages=21]{supp.pdf}}
\AtEndDocument{\includepdf[pages=22]{supp.pdf}}
\AtEndDocument{\includepdf[pages=23]{supp.pdf}}
\AtEndDocument{\includepdf[pages=24]{supp.pdf}}
\AtEndDocument{\includepdf[pages=25]{supp.pdf}}
\AtEndDocument{\includepdf[pages=26]{supp.pdf}}
\AtEndDocument{\includepdf[pages=27]{supp.pdf}}
\AtEndDocument{\includepdf[pages=28]{supp.pdf}}
\AtEndDocument{\includepdf[pages=29]{supp.pdf}}
\AtEndDocument{\includepdf[pages=30]{supp.pdf}}
\AtEndDocument{\includepdf[pages=31]{supp.pdf}}
\AtEndDocument{\includepdf[pages=32]{supp.pdf}}
\AtEndDocument{\includepdf[pages=33]{supp.pdf}}
\AtEndDocument{\includepdf[pages=34]{supp.pdf}}
\AtEndDocument{\includepdf[pages=35]{supp.pdf}}
\AtEndDocument{\includepdf[pages=36]{supp.pdf}}

\end{document}